# Effect of tilt and twist angles on the cleavage crack propagation

Vipul Jain[1], AryaChatterjee[2], S. Patra[3,4], D. Chakrabarti[3], A. Ghosh[1]*

[1]Discipline of Metallurgy Engineering and Materials Science, IIT Indore, Simrol, M.P., India.
[2]School of Engineering, Brown University, USA.
[3]Department of Metallurgical and Materials Engineering, IIT Kharagpur, Kharagpur, W.B., India.
[4]Jindal Stainless Limited, Hisar, H.R.,India.
*Corresponding author

**Abstract:**

We report the mechanism of cleavage crack deviation at grain boundaries. A fracture mechanics based model has been developed to combine the effect of tilt and twist angles between cleavage planes on the reduction in normalized energy release rate. Furthermore, electron back-scatter diffraction analysis on cleavage fracture surface has been carried out. The microstructural features associated with the presence of tilt and twist angle at the grain boundary have been identified and found to be consistent with the prediction from the analytical model.

Keywords: Fracture, Cleavage facet, Electron back-scatter diffraction, Grain boundary, Tilt angle, Twist angle,

**1.0  Introduction:**

The phenomenon of catastrophic cleavage fracture in body-centered cubic (BCC) material limits its application at low temperature and high strain rate conditions. The resistance to cleavage fracture in BCC material can be improved by reducing grain size and increasing grain boundary misorientation angle (GBMA). The effect of reduced grain size on cleavage fracture resistance is well studied and understood[1–3]. Comparatively, less attention has been paid on how GBMA influences crack propagation. Probably, the reason lies among difficulties associated with Electron back-scatter diffraction (EBSD) analysis on the fracture surface. Therefore, a few studies have been conducted to correlate cleavage fracture surface morphology with grain orientation[4–6]. One of the pioneering works done by Bhattacharjee et al. [4] involves automatic-EBSD scan over the fracture surface of ferritic steel. They concluded that angular misorientation up to 12⁰ (angle-axis pair) can exist within a single cleavage facet which defines the effective-grain size[7,8].They also confirmed that cleavage crack in BCC ferrite propagates along {001} crystal planes. The justification provided for 12º threshold GBMA was 5% reduction in crack driving force. However, the reduction in crack driving force depends on tilt and twist angle rather than GBMA. Furthermore, the crystallographic nature of the cleavage plane can be obtained from inverse pole figure, if the cleavage facet is parallel to the macroscopic fracture plane, which is improbable mostly. A stereological correction before EBSD scan is necessary to characterize the crystallographic cleavage plane, as reported by Randle et al.[5]. EBSD analysis on the fracture surface is very challenging and there exists a high-risk of wrong interpretation of results. Moreover, a disagreement regarding the correctness of EBSD data on the fracture surface persists since the facet deviates from an ideal 70° tilt condition.

In our previous work[9], it has been shown that the cleavage crack deviation depends on the angular difference between adjacent cleavage planes along the propagating crack, and the experimental observations of crack deviations were correlated with that angle considering the angle of projection on the observed surface. However, the role of these angular differences on the cleavage fracture surface morphology, and their effect on imposing a barrier to cleavage crack propagation have not been studied thoroughly till date. Understanding the role of grain boundary (GB) nature on crack deviation is crucial, and recently it has been reported that GB nature severely affects the overall impact toughness[10,11]. The present work does not aim to correlate GB character with the overall property, rather focuses on how the tilt and twist angles of a GB influence the crack-driving force associated with the formation of different features at GB.

## 2.0    Experimental Details:

Standard Transverse-Longitudinal oriented Charpy impact sample, prepared from rolled and annealed plate of ferritic steel with a large grain size (60 μm), was subjected to impact loading at -196ºC. To study the fracture surface of those specimens, they were cut along the plane parallel to the macroscopic fracture plane with roughly 2 mm thickness using a slow speed diamond cutter, followed by ultrasonic cleaning and drying. Special caution has been taken to keep the flat cut surface exactly perpendicular to transverse direction (TD), as shown in **Fig. 1(a)**. Neither mechanical nor electro-polishing has been carried out on the fracture surface. Sample was placed on the 70° pre-tilted holder. EBSD analysis was carried out using Aztec system (Oxford Instruments, Abingdon, Oxfordshire, UK) attached to Zeiss

Auriga Compact scanning electron microscope (SEM). Although the actual fracture surface is not at 70° tilt condition and tilting angle varies at a great extent from facet to facet, the condition of 70° tilting about the reference axis (i.e. TD) remains constant for all the cleavage facets. Therefore, the orientation obtained by EBSD analysis from different regions of the fracture surface is correct irrespective of their actual tilt angle. Instead, because of the arbitrary inclination of the neighboring cleavage facets, the Kikuchi pattern generated from a facet could not reach up to the camera. As a consequence, the Kikuchi pattern recorded on the camera used to have some partial obstacle and may not be suitable for automatic analysis. So, in the present study, EBSD analysis has been carried out manually, point by point. In addition, the indexing of the Kikuchi pattern has been restricted to the obstructer free portion of the recorded pattern which was chosen manually. The orientation data is collected from at least five neighbouring points for a single facet, and the GBMA between different points of a single facet is always found to be less than 0.5°.

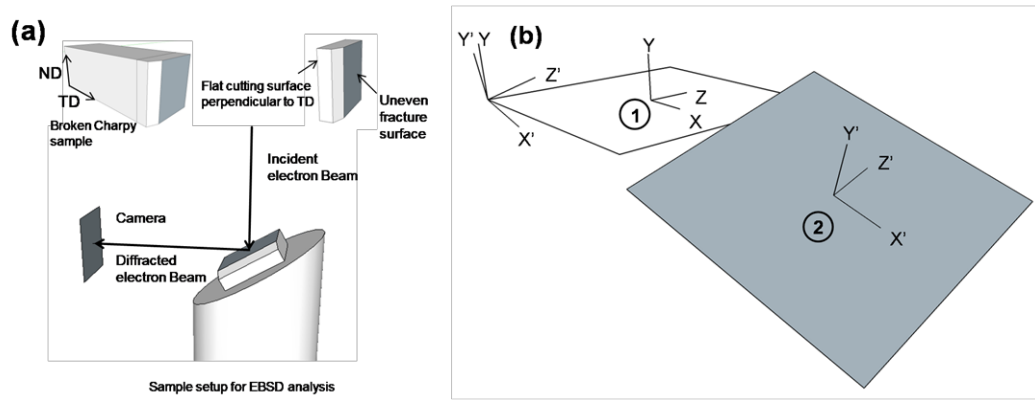

**Figure 1:** (a) Details of sample preparation and setup configuration for EBSD analysis (b) Model for out of plane crack extension with tilt and twist angle.

### 3.0  Analytical model for energy release rate:

Cleavage crack follows a path of specific crystallographic planes during its propagation and deviates at GB depending upon the neighbouring grains orientations. Some earlier studies focused on the individual effect of either tilt or twist angle on crack driving[12,13]. Bhattacharjee et al.[14] followed a simplified approach to estimate the reduction in crack driving by considering GBMA (i.e. angle-axis pair) instead of tilt twist angles which can lead to huge deviation. It will be more realistic and accurate to consider the combined effect of tilt and twist angle. Simultaneously, understanding the reduction in crack driving at GB is important from the perspective of defining effective-grain size[7,8,14]. In the present study, an attempt has been made to develop an analytical model to estimate the reduction in normalized energy release rate (ΔG) as a consequence of the combined tilt-twist angle, originated from crack deviation at GB.

The Schematic diagram, presented in **Fig. 1(b)** shows two cleavage planes at arbitrary tilt-twist angles. Let us assume, the stress intensity factors for local mode-I, II, and III are **$k_I$, $k_{II}$, $k_{III}$**, respectively. The remote mode-I stress intensity factor is **$K_I$**, normalized stress tensor for remote mode-I loading is **$S_\sigma$**, and the normalized stress tensor associated with the tilted and twisted facet is $S'_\sigma$.

The components of normalized stress tensor on plane 1($S_\sigma$) under mode-I loading in-plane strain condition are as follows[15]:

$$\sigma_{xx} = \cos\left(\frac{\alpha}{2}\right)\left[1 - \sin\left(\frac{\alpha}{2}\right)\sin\left(\frac{3\alpha}{2}\right)\right]$$

$$\sigma_{yy} = \cos\left(\frac{\alpha}{2}\right)\left[1 + \sin\left(\frac{\alpha}{2}\right)\sin\left(\frac{3\alpha}{2}\right)\right]$$

$$\sigma_{zz} = \upsilon(\sigma_{xx} + \sigma_{yy})$$

$$\tau_{xy} = \cos\left(\frac{\alpha}{2}\right)\sin\left(\frac{\alpha}{2}\right)\cos\left(\frac{3\alpha}{2}\right)$$

$$\tau_{xz} = \tau_{yz} = 0$$

Where, $\alpha$ is the tilt angle and ν, Poisson's ratio is assumed to be 0.33[16].

Depending upon the orientation of the neighbouring crystal, GB plane and cleavage plane, one can obtain the tilt (α) and twist (φ) angle of the deviated crack from the following approach proposed by King et al.[17,18]:

$$\varphi = \cos^{-1}[(P_{gb} \times P_{c1}) \bullet (P_{gb} \times P_{c2})]$$

Where $P_{gb}$ is the unit normal to GB plane and $P_{C1}$ and $P_{C2}$ are unit normal to the cleavage plane of two neighbouring crystal-1 and 2, **Fig 1b** in the sample reference frame. $P_{C1}$ and $P_{C2}$ can be obtained by incorporating the orientation of the crystals.

Now, tilt angle can be obtained after eliminating the twisting effect as follows:

$$P'_{c2} = Rot \times P_{c2};$$

Here, **Rot** represents the rotation about $P_{gb}$ axis with the twist angle φ. Hence, the tilt angle (α) can be measured by the following equation:

$$\alpha = \cos^{-1}[(P_{gb} \times P_{c1}) \bullet (P_{gb} \times P'_{c2})]$$

The rotation matrices associated with the tilt angle (α) and twist angle (φ) are as follows:

$$R_{tilt(z)} = \begin{bmatrix} \cos\alpha & \sin\alpha & 0 \\ -\sin\alpha & \cos\alpha & 0 \\ 0 & 0 & 1 \end{bmatrix}$$

$$R_{twist(x)} = \begin{bmatrix} 1 & 0 & 0 \\ 0 & \cos\phi & \sin\phi \\ 0 & -\sin\phi & \cos\phi \end{bmatrix}$$

The resultant rotational matrix ($R_t$) due tiltand twist can be given by:

$$R_t = R_{twist(x)} \cdot R_{tilt(z)}$$

Therefore, the normalized state of stress on new plane-2, **Fig 1b** due to remote mode-I loading can be obtained by the coordinate transformation as follows:

$$S'_\sigma = R_t \cdot S_\sigma \cdot R^T_t$$

Now, the ratio of local to remote stress intensity factor associated with facet-2, **Fig. 1b** can be expressed by different components of the normalized stress tensor ($S'_\sigma$) as follows [12,13]:

$$\frac{k_I}{K_I} \approx \sigma_{y'y'} : S'_\sigma$$

$$\frac{k_{II}}{K_I} \approx \tau_{x'y'} : S'_\sigma$$

$$\frac{k_{III}}{K_I} \approx \tau_{y'z'} : S'_\sigma$$

Finally, the reduction in normalized energy release rate (ΔG) can be given by:

$$\Delta G = 1 - \left[\left(\frac{k_I}{K_I}\right)^2 + \left(\frac{k_{II}}{K_I}\right)^2 + \left(\frac{k_{III}}{K_I}\right)^2\right]$$

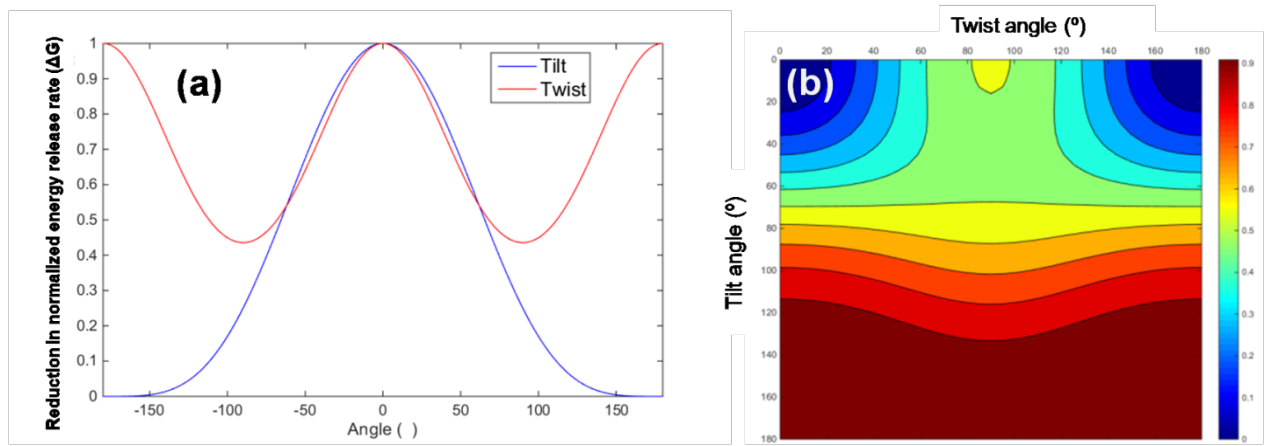

**Figure 2:** Reduction in normalized energy release rate (a) for Individual tilt and twist angle and (b) for a combined tilt-twist angle.

The predicted values of ΔG for different tilt and twist angles are shown in **Fig. 2,** following the above-mentioned model. ΔG value for individual tilt and twist angle is almost similar within the range of ±60º, **Fig 2(a)**. Beyond this point (i.e. ± 60º), ΔG is more for tilt than for twist angle, **Fig 2(a)**. The combined effect of tilt and twist angle on ΔG is represented by the colour code in **Fig. 2(b)**. It is evident that tilt angle is more effective in reducing crack driving as compared to the twist angle.

## 4.0  Experimental results and discussion:

**Fig. 3** shows the fractographic region chosen for EBSD analysis. The mode of fracture is completely transgranular cleavage as the impact test was carried out at -196 °C. The fracture surface is flat and consists of several cleavage facets. Different types of boundaries between the facets are observed and indicated by red arrows.

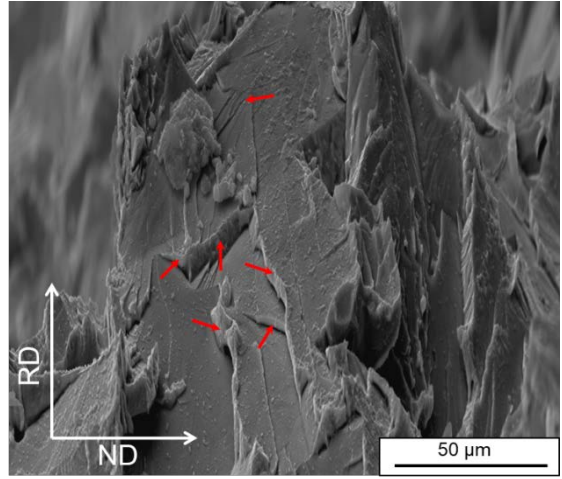

**Figure 3**: SEM image of the fracture surface showing the presence of cleavage facets and different types of boundaries.

EBSD analysis has been carried out at different regions of the fracture surface to understand the role of relative misorientation between the neighbouring crystals on their boundary morphology. It is well established that the cleavage fracture occurs along {001} plane in ferritic and martensitic steel[4,5,7,8,14], therefore {001} planes are considered to be the cleavage plane in the present calculation. Now, there are three possible cleavage planes from each crystal therefore, there are nine possible combinations of tilt and twist angles between the neighboring grains. Now, depending on the actual tilt-twist angle, cleavage crack loses its driving force as per the details discussed in **section 3**. In the present study, it is proposed that a cleavage crack will follow the particular cleavage plane which provides minimum possible ΔG. This hypothesis will evaluated considering three different cases as follows.

**Figure 4:** (a, d, g) The SEM fractographs showing cleavage facets separated by different types of boundaries;(b,c,e,f,h,i) The Kikuchi pattern obtained from the locations 1,2,3,4,5 and 6, respectively, marked in the fractographs (a, d, g).

**Table 1**: Eulers' angle with the mean angular deviation (MAD) and misorientation angle(MA) as obtained from the point EBSD analysis on the cleavage facet.

| Region | Orientation in Euler angle (°) | | | MAD (°) | MA (°) |
|---|---|---|---|---|---|
| | $\varphi_1$ | $\varphi$ | $\varphi_2$ | | |
| 1 | 7.88 | 82.89 | 56.21 | 0.45 | 7.83 |
| 2 | 4.26 | 84.07 | 63.48 | 0.63 | |
| 3 | 128.41 | 25.89 | 47.96 | 0.51 | 14.72 |
| 4 | 9.92 | 80.36 | 60.16 | 0.56 | |
| 5 | 69.99 | 74.9 | 5.88 | 0.34 | 41.6 |
| 6 | 125.87 | 24.3 | 48.5 | 0.75 | |

**Case 1**:

The cleavage facets of interest are indicated by region **1** and **2** in **Fig. 4(a)**. The boundary between these regions is a typical example of tilt boundary. Kikuchi patterns collected from these regions corresponding to two neighbouring facet, as shown in **Fig. 4(b,c)**. The Kikuchi patterns in **Fig. 4(a)** and **(c)** are not suitable for automatic analysis because it contains some partial obstacle which appears as a black region at the bottom of the figure. Therefore, the zone axis identification process is restricted in the region surrounded by a white circle, Fig. **Fig. 4(b,c)**. The same procedure is followed for other regions. The orientation obtained by EBSD analysis on region **1** and **2** is given in **Table 1**. The MA between them is 7.83°. Conventionally, any MA less than 10°-15° is considered to be low angle boundary, corresponding to a single facet[8,9,14]. Interestingly, this result conflicts with the conventional understanding. Since Charpy samples were prepared from the rolled and annealed plate, most of GB are perpendicular to ND of the rolling plate[10,17]. The nine possible combinations of tilt and twist angles considering the orientations of **regions 1** and **2** are reported in **Table 2** and ΔG for all possible combinations are shown in **Table 3.** The twist angle is minimum (i.e. 1.8º) for the combination of (010)-(010) cleavage plane, while the combination of (001)-(001) cleavage planes have minimum possible tilt angle (i.e. 1.2º).Evidently, the three combinations of cleavage planes, such as (100)-(100), (010)-(010) and (001)-(001) result in ΔG within 1%. In the present study it was not possible to measure the tilt angle experimentally. However, it appears to be very small between the facets, **Fig. 4(b).**The experimental observation is also consistent with predicted values of the proposed analytical model, **Table 2**.

**Table 2**: Calculated values of tilt and twist angles for the different combinations of cleavage planes.

| Case1 (Grain boundary Perpendicular to ND) | | | | Case2 (Grain boundary Perpendicular to ND) | | | | Case3 (Grain boundary Perpendicular to RD) | | | |
|---|---|---|---|---|---|---|---|---|---|---|---|
| Twist angle (°) | | | | Twist angle (°) | | | | Twist angle (°) | | | |
| Cleavage Plane | 100 | 010 | 001 | Cleavage Plane | 100 | 010 | 001 | Cleavage Plane | 100 | 010 | 001 |
| 100 | 2.4 | 17.1 | 75.9 | 100 | 32.8 | 11.1 | 73.4 | 100 | 60.1 | 21.8 | 69.2 |
| 010 | 12.8 | 1.8 | 88.9 | 010 | 63.2 | 84.9 | 10.6 | 010 | 24.4 | 73.7 | 15.3 |
| 001 | 81.9 | 83.4 | 3.6 | 001 | 12.2 | 34.0 | 61.5 | 001 | 75.7 | 22.4 | 66.6 |
| Tilt angle (°) | | | | Tilt angle (°) | | | | Tilt angle (°) | | | |
| Cleavage Plane | 100 | 010 | 001 | Cleavage Plane | 100 | 010 | 001 | Cleavage Plane | 100 | 010 | 001 |
| 100 | 7.3 | 38.2 | 49.6 | 100 | 39.9 | 22.7 | 28.6 | 100 | 51.9 | 15.5 | 37.9 |
| 010 | 29.4 | 7.1 | 39.4 | 010 | 75.8 | 12.4 | 21.8 | 010 | 54.1 | 19.1 | 29.2 |
| 001 | 70.0 | 19.3 | 1.2 | 001 | 5.3 | 54.3 | 54.5 | 001 | 43.2 | 62.2 | 45.6 |

**Table 3**: Reduction in normalized energy release rate for different combination cleavage planes.

| Case1 (Grain boundary Perpendicular to ND) | | | | Case2 (Grain boundary Perpendicular to ND) | | | | Case3 (Grain boundary Perpendicular to RD) | | | |
|---|---|---|---|---|---|---|---|---|---|---|---|
| Fraction reduction in crack driving energy | | | | Fraction reduction in crack driving energy | | | | Fraction reduction in crack driving energy | | | |
| Cleavage Plane | 100 | 010 | 001 | Cleavage Plane | 100 | 010 | 001 | Cleavage Plane | 100 | 010 | 001 |
| 100 | 0.01 | 0.23 | 0.51 | 100 | 0.32 | 0.10 | 0.51 | 100 | 0.48 | 0.12 | 0.49 |
| 010 | 0.15 | 0.01 | 0.52 | 010 | 0.59 | 0.55 | 0.09 | 010 | 0.40 | 0.52 | 0.16 |
| 001 | 0.56 | 0.54 | 0.01 | 001 | 0.03 | 0.42 | 0.49 | 001 | 0.51 | 0.47 | 0.49 |

**Case 2**:

A ridge zone is observed between the two facets at **regions3** and **4**, **Fig. 4(d)**. Kikuchi patterns corresponding to these regions are shown in **Fig. 4(e)** and **(f)**, respectively. The GBMA between two facets is around 15°, **Table 1**. Here the lowest possible twist angle is 11.1° between (100)-(010) cleavage planes, while the minimum tilt angle (5.3°) is between (001)-(100) cleavage planes, **Table 2**. Also, the minimum possible ΔG is 3% which is for the combination of (001)-(100) cleavage planes, **Table 3** and the corresponding tilt and twist angles are 5.3º and 11.2º respectively. As per the analytical model, boundary contains both tilt and twist angle between **regions 3** and **4**. Following the prediction, a microstructural feature comprising a ridge zone which is expected to provide a significant barrier [10,11], is experimentally observed at the facet boundary.

**Case 3**:

A classical step formation is observed at the boundary between **regions 5** and **6**,**Fig.4(g).** The Kikuchi patterns corresponding to **regions 5** and **6** are shown in **Fig 4(h)** and **(i)** respectively. The boundary between these two facets is perpendicular to RD, unlike the other cases. GBMA between the two regions is 41.6°, **Table 1**. Minimum possible twist angle is around 15.3°, and the corresponding tilt angle is 15.5°, **Table 2**. Also, the minimum possible ΔG is 12% which is for the combination of (100)-(010) cleavage plane, **Table 3**. Therefore, the analytical model indicates a high twist angle at the boundary which is consistent with observation made at the facet boundary, **Fig. 4(g)** and **Table 2**. The boundary between **region 5** and **6** is associated with step formation as consequence of the high twist angle between them and is expected to impose a strong barrier to the cleavage crack propagation.

In summary for all three cases, the predicted ΔG is found to be minimum for the combination of cleavage plane where the tilt angle is minimum. It may be because the tilt angle has more influence on ΔG. But unlike tilt angle, twist angle at facet boundary creates an additional surface which is not considered in the proposed model. It (The consideration of twist angle) may be may be necessary for better understanding and will be addressed in our future work.

## 5.0 Conclusions

The following conclusions can be derived from the above study:

(a) The actual crack is found to propagate along the cleavage planes which possess the lowest possible ΔG, as predicted by the analytical model.
(b) The tilt angle is found to have more influence on ΔG as compared to the twist angle which causes step formation at GB.

**Acknowledgement:**

Dr. A. Ghosh is thankful to SERB-DST, Govt. of India for providing financial support (Grant no.IFA17-ENG193).

**References**

[1]   J.F. Knott, Fundamentals of Fracture Mechanics, Gruppo Italiano Frattura, 1973.

[2]   A. Ghosh, A. Ray, D. Chakrabarti, C.L. Davis, Mater. Sci. Eng. A 561 (2013) 126–135.

[3]   D. Chakrabarti, M. Strangwood, C. Davis, Metall. Mater. Trans. A 40 (2009) 780–795.

[4]   D. Bhattacharjee, C.L. Davis, Scr. Mater. 47 (2002) 825–831.

[5]   V. Randle, P. Davies, Mater. Sci. Technol. 21 (2005) 1275–1281.

[6]   D.C. Slavik, J.A. Wert, R.P. Gangloff, J. Mater. Res. 8 (1993) 2482–2491.

[7]   N.J. Kim, A.H. Nakagawa, A.H. Nakagawa, Mater. Sci. Eng. 83 (1986) 145–149.

[8]   M.-C. Kim, Y. Jun Oh, J. Hwa Hong, Scr. Mater. 43 (2000) 205–211.

[9]   A. Ghosh, S. Kundu, D. Chakrabarti, Scr. Mater. 81 (2014) 8–11.

[10]  A. Ghosh, S. Patra, A. Chatterjee, D. Chakrabarti, Metall. Mater. Trans. A 47 (2016) 2755–2772.

[11]  A. Chatterjee, A. Ghosh, A. Moitra, A.K. Bhaduri, R. Mitra, D. Chakrabarti, Int. J. Plast. 104 (2018) 104–133.

[12]  B. Lawn, T.R. Wilshaw, Fracture of Brittle Solids, Cambridge university press, 1993.

[13]  K.T. Faber, A.G. Evans, Acta Metall. 31 (1983) 565–576.

[14]  D. Bhattacharjee, J.F. Knott, C.L. Davis, Metall. Mater. Trans. A 35 (2004) 121–130.

[15]  T.L. Anderson, Fracture Mechanics: Fundamentals and Applications, CRC press, 2017.

[16]  W. KÖSTER, H. FRANZ, Metall. Rev. 6 (1961) 1–56.

[17] A. King, W. Ludwig, M. Herbig, J.-Y. Buffière, A.A. Khan, N. Stevens, T.J. Marrow, Acta Mater. 59 (2011) 6761–6771.

[18] T. Zhai, A.J. Wilkinson, J.W. Martin, Acta Mater. 48 (2000) 4917–4927.